\documentclass[twocolumn]{emulateapj}

\usepackage{amssymb,graphicx}
\usepackage{epsfig}
\usepackage{psfrag}
\usepackage[usenames]{color}
\usepackage{multirow}
\usepackage{rotating}

\newcommand{\eqref}[1]{(\ref{#1})}

\shorttitle{GRMHD simulations of binary BHNS: the jet emerges}
\shortauthors{Paschalidis, Ruiz \& Shapiro}

\begin{document}

\title{Relativistic simulations of black hole--neutron star
  coalescence: the jet emerges} 

\author{Vasileios Paschalidis${}^{1,2}$, Milton Ruiz${}^{2}$, Stuart
  L. Shapiro${}^{2,3}$} \affil{ ${}^1$Department of Physics, Princeton
  University, Princeton, NJ~08544\\ ${}^2$Department of Physics,
  University of Illinois at Urbana-Champaign, Urbana, IL~61801
  \\ ${}^3$Department of Astronomy \& NCSA, University of Illinois at
  Urbana-Champaign, Urbana, IL~61801 }

\begin{abstract}
We perform magnetohydrodynamic simulations in full general relativity
(GRMHD) of a binary black hole--neutron star on a quasicircular orbit
that undergoes merger. The binary mass ratio is $3:1$, the black hole
initial spin parameter $a/m=0.75$ ($m$ is the black hole Christodoulou
mass) aligned with the orbital angular momentum, and the neutron star
is an irrotational $\Gamma=2$ polytrope. About two orbits prior to
merger (at time $t=t_B$), we seed the neutron star with a dynamically
weak interior dipole magnetic field that extends into the stellar
exterior. At $t=t_B$ the exterior has a low-density atmosphere with
constant plasma parameter $\beta\equiv P_{\rm gas}/P_{\rm
  mag}$. Varying $\beta$ at $t_B$ in the exterior from $0.1$ to
$0.01$, we find that at a time $\sim 4000{\rm M} \sim 100(M_{\rm
  NS}/1.4M_\odot)$ms [M is the total (ADM) mass] following the onset
of accretion of tidally disrupted debris, magnetic winding above the
remnant black hole poles builds up the magnetic field sufficiently to
launch a mildly relativistic, collimated outflow -- an incipient
jet. The duration of the accretion and the lifetime of the jet is
$\Delta t\sim 0.5(M_{\rm NS}/1.4M_\odot)$s.  Our simulations furnish
the first explicit examples in GRMHD which show that a jet can emerge
following a black hole - neutron star merger.
\end{abstract}

\maketitle

\section{Introduction}

Black hole--neutron star (BHNS) binaries are promising sources for
detectable gravitational waves (GWs) by ground-based laser
interferometers such as aLIGO~\citep{LIGO2}, VIRGO~\citep{VIRGO1},
GEO~\citep{GEO}, and KAGRA~\citep{LCGT}. Moreover, mergers of BHNSs
have been suggested as central engines that power short-hard gamma ray
bursts (sGRBs); see, e.g.,
\citep{Meszaros:2006rc,LeeRamirezRuiz2007}. The GW signal from the
inspiral and merger, the amount and composition of ejecta from BHNSs
and effects of different equations of state (EOS), have been explored
in full general relativity (GR)~\citep[see,
  e.g.][]{UIUC_BHNS__BH_SPIN_PAPER,
  Duez:2009yy,st11,East2012,Lovelace:2013vma, Lackey:2013axa,
  Kyutoku:2013wxa, Deaton:2013sla,Pannarale:2013uoa, Tanaka:2013ixa,
  Foucart:2014nda}.  Studies of magnetized BHNS mergers in full GR
magnetohydrodynamics (MHD) also have been carried out
\citep{cabllmn10,Etienne:2011ea,Etienne:2012te}.
Magnetospherically-powered precursor EM signals from BHNS systems have
been simulated as well~\citep[][]{Paschalidis:2013jsa}.

While these studies have made great progress, to date, there exists no
self-consistent calculation in full GR that starts from the late BHNS
inspiral and demonstrates that jets can be launched from BHNSs. This
step is crucial to establishing BHNSs as viable central engines for
sGRBs and solidifying their role as multimessenger systems.

It is known that if the initial NS has a B field confined to its
interior, the B-field lines following the BHNS merger are wound into
an almost purely toroidal configuration and no jets are launched
\citep{Etienne:2011ea,Etienne:2012te}. The existence of near purely
toroidal B fields in the remnant disk explains why jets cannot be
launched magnetically: for a magnetized accretion disk with B fields
initially confined to the disk interior, a jet is launched and
supported only if these initial seed fields have poloidal components
with a consistent sign in the vertical direction
\citep{GRMHD_Jets_Req_Strong_Pol_fields}. Applying this principle, we
demonstrated that by artificially seeding with a purely poloidal B
field the remnant disk formed in a hydrodynamic simulation of a BHNS
merger, an incipient jet is indeed launched~\citep{Etienne:2012te}.
Thus, under ``right conditions,'' jets can be magnetically launched
from BHNSs. However, identifying the initial configuration prior to
tidal disruption that leads to these ``right conditions'' remains
elusive. We note that there exists a calculation for binary NSs that
reports jet formation~\citep{ML2011}, but a more recent
high-resolution study~\citep{Kiuchi:2014hja} suggests otherwise.

\begin{figure*}
  \centering
  \includegraphics[width=0.33\textwidth]{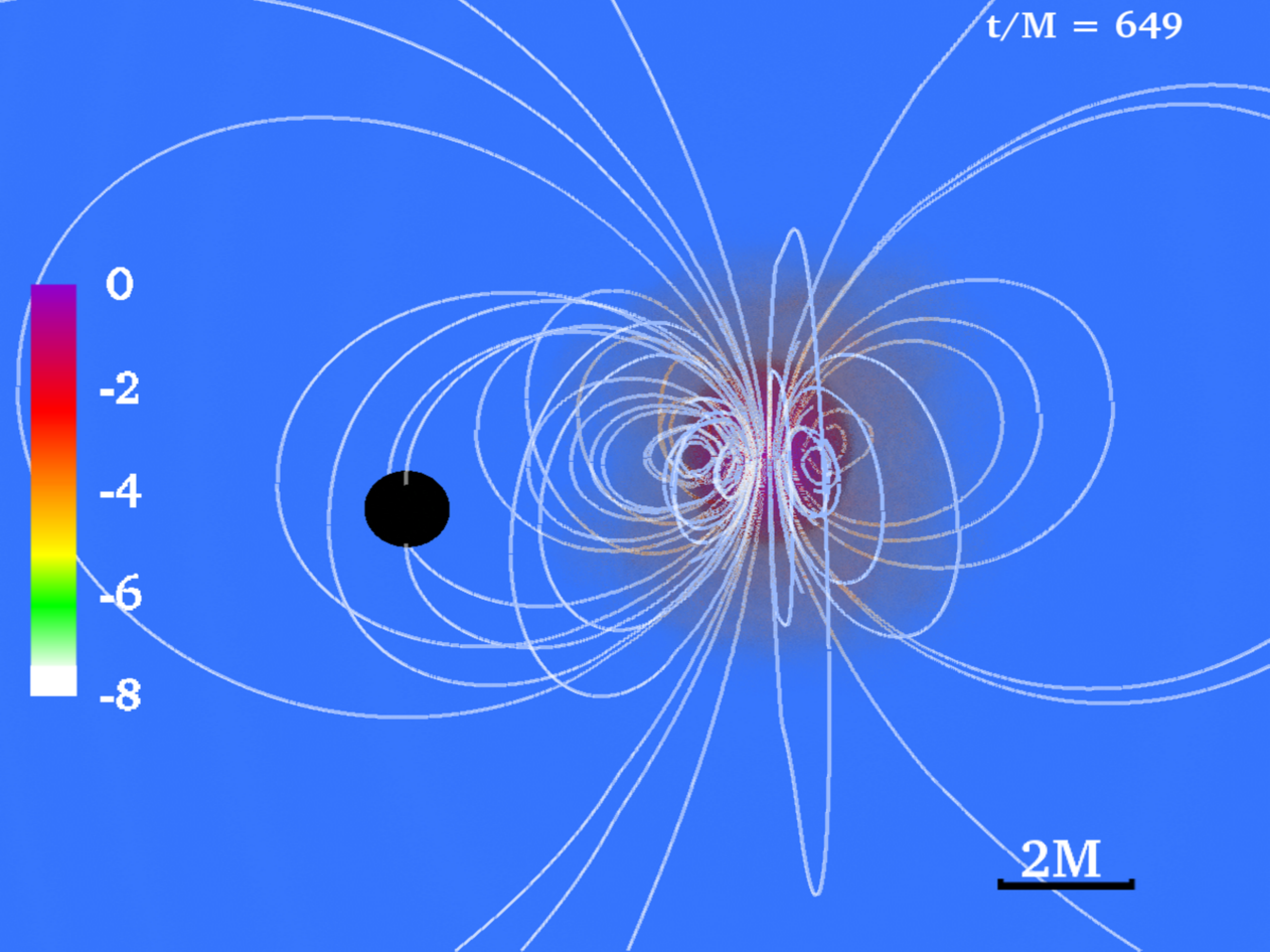}
  \includegraphics[width=0.33\textwidth]{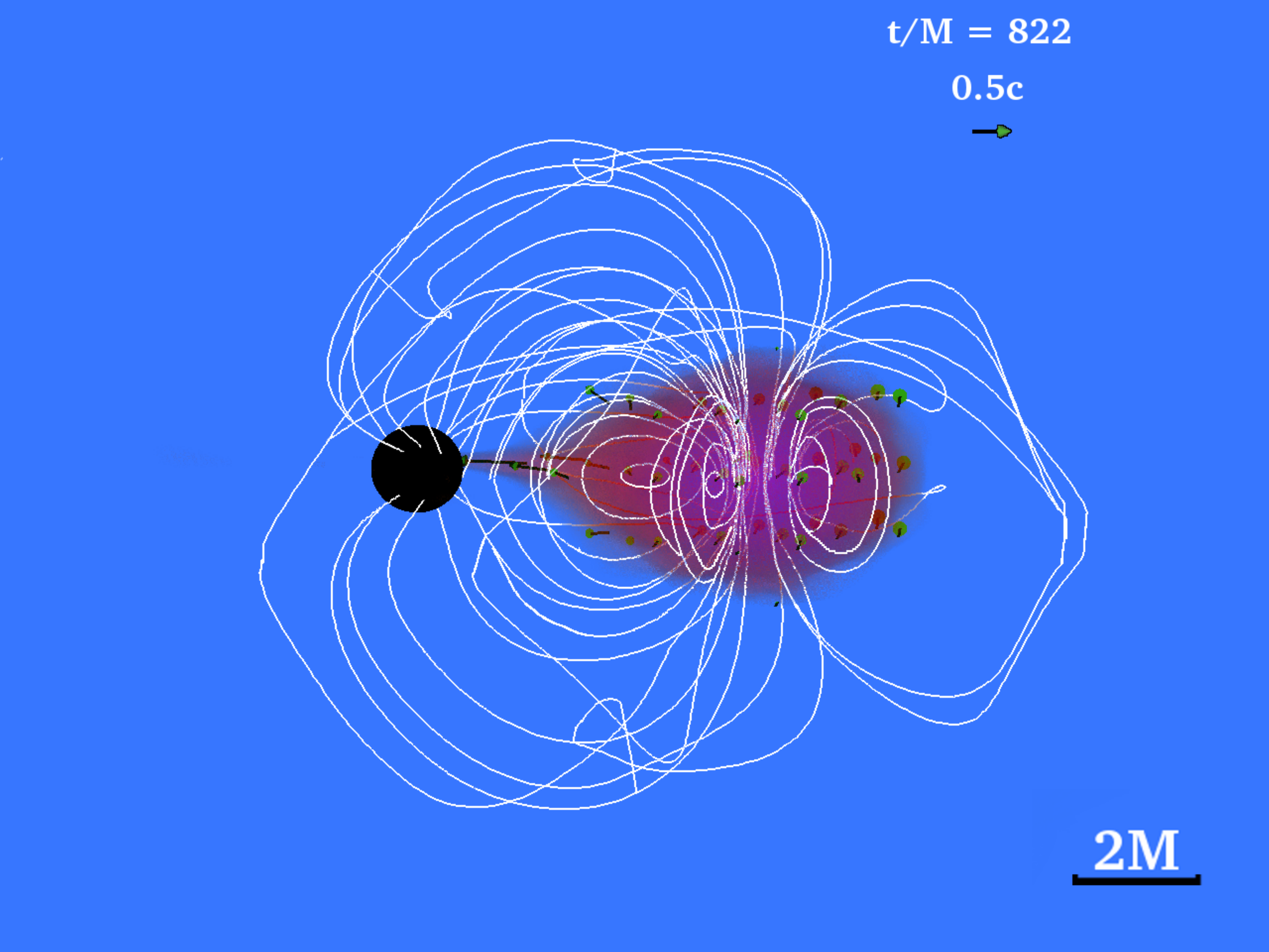}
  \includegraphics[width=0.33\textwidth]{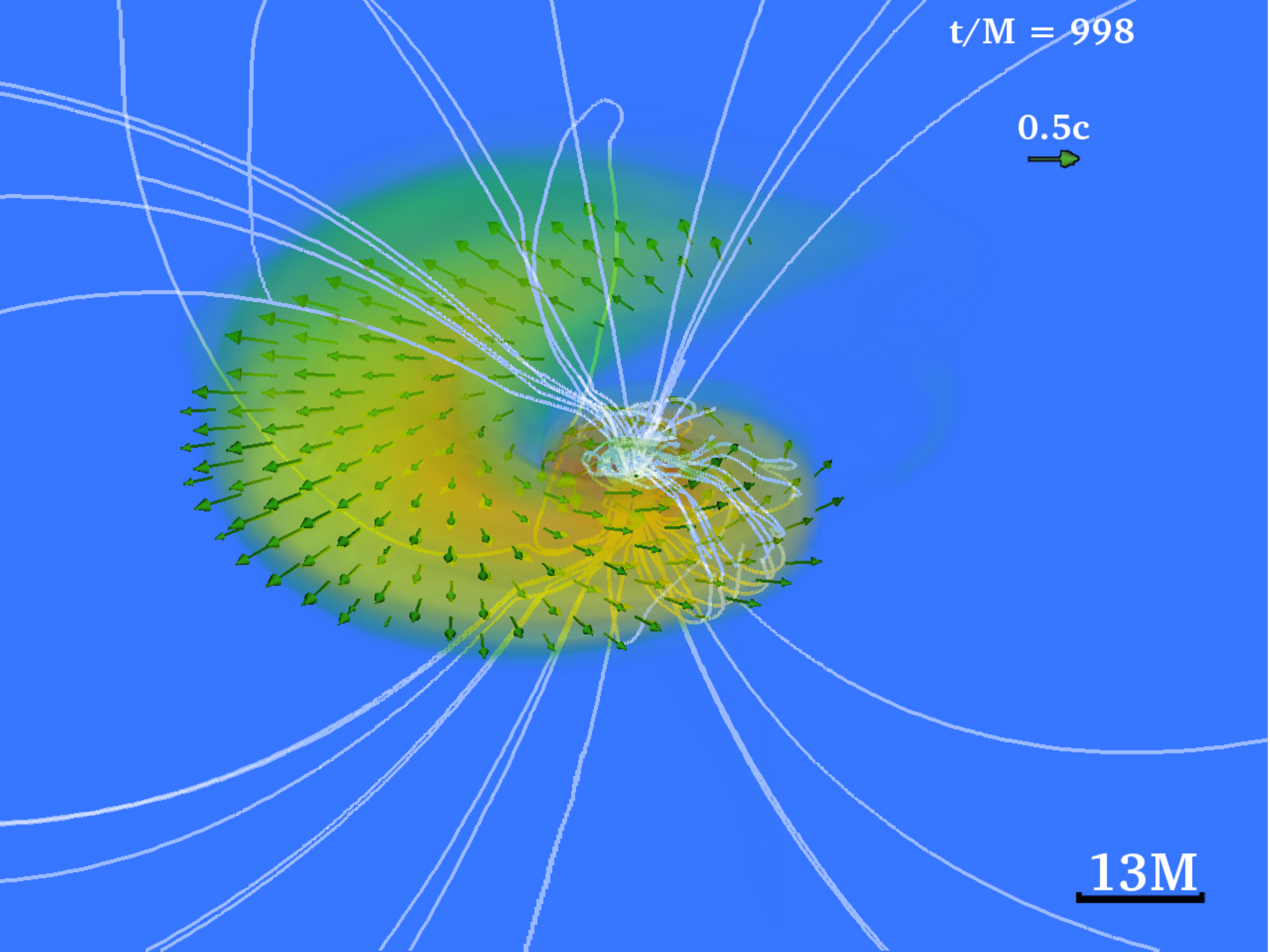}
  \includegraphics[width=0.33\textwidth]{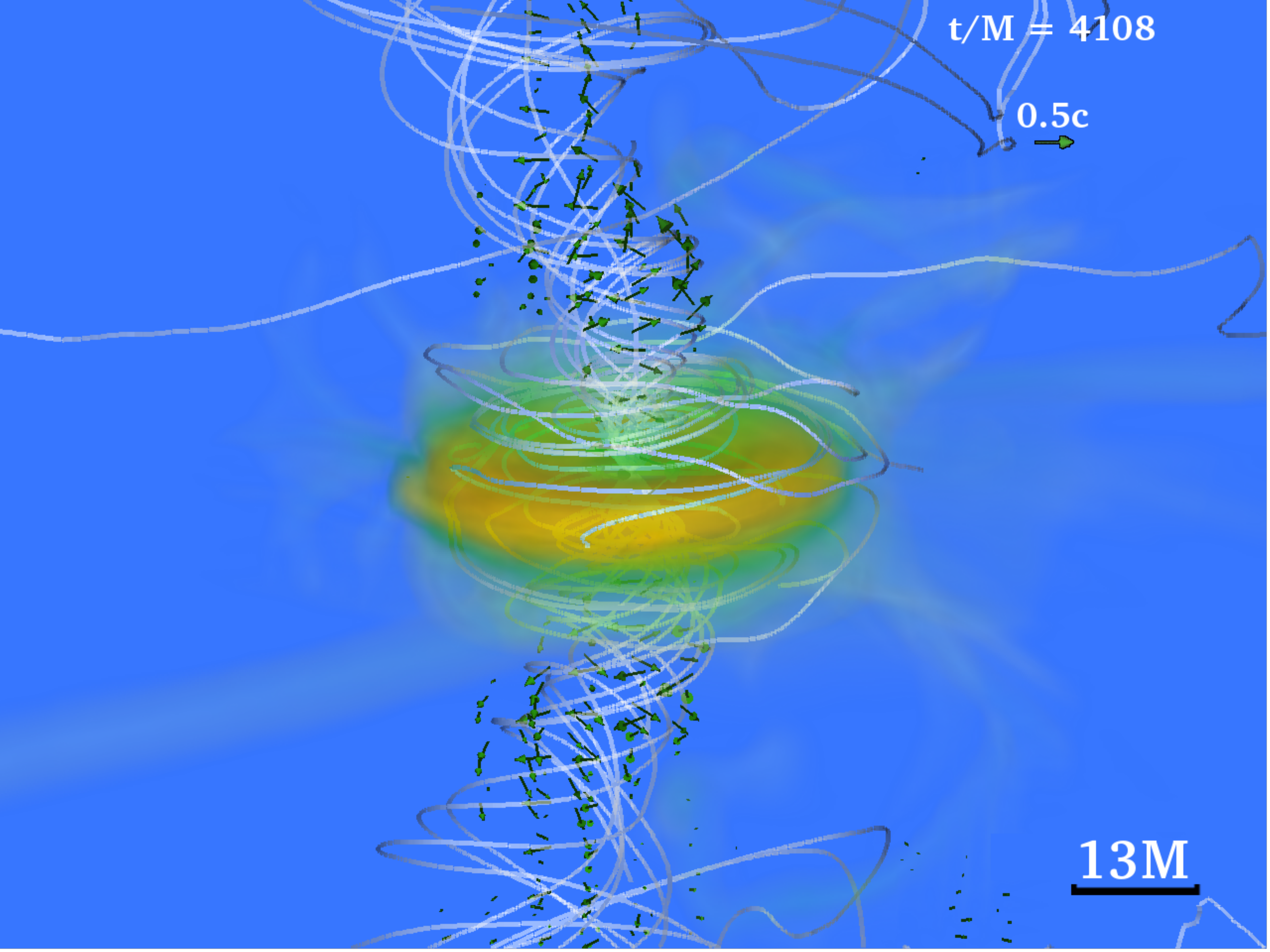}
  \includegraphics[width=0.33\textwidth]{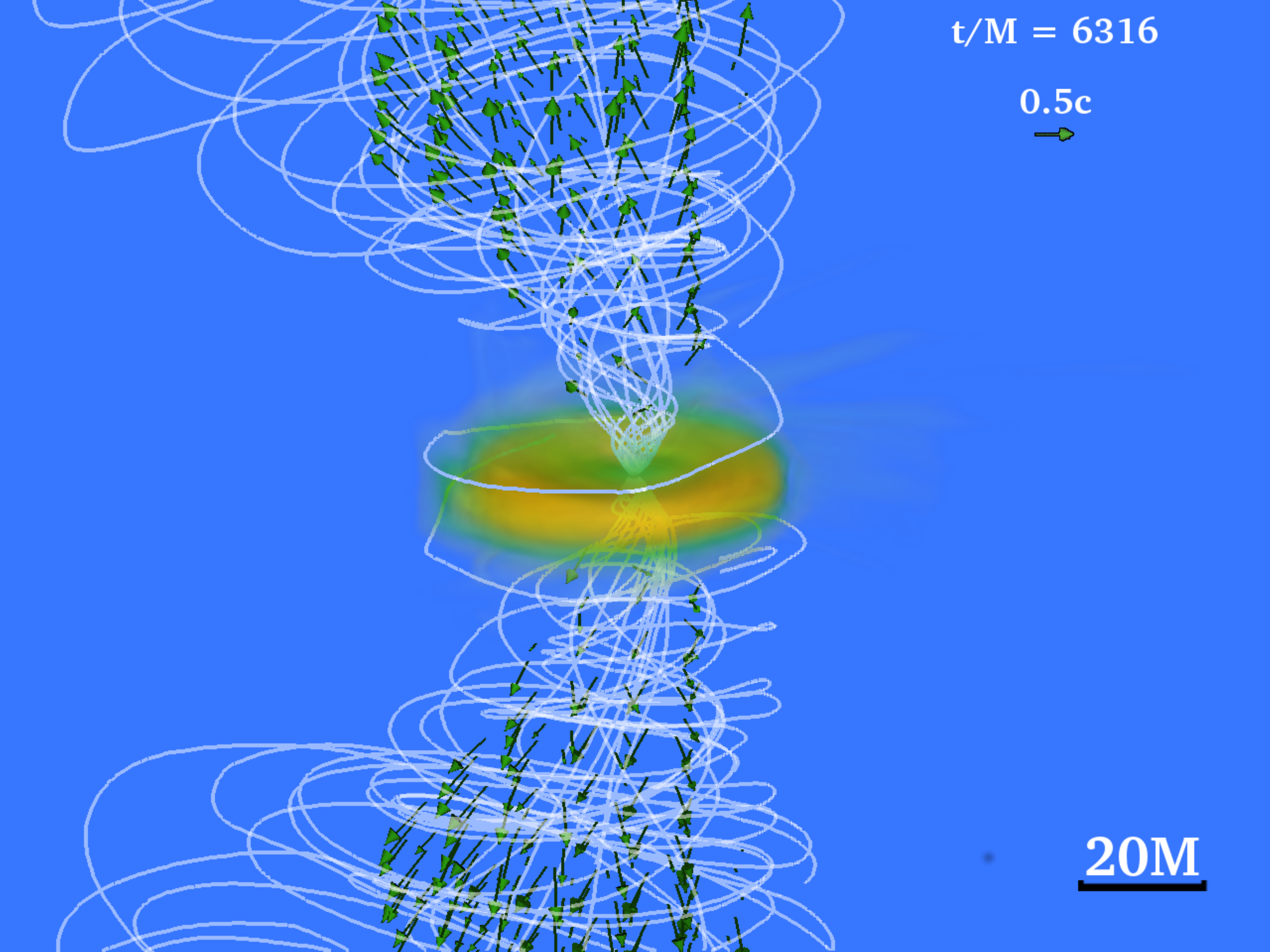}
  \includegraphics[width=0.33\textwidth]{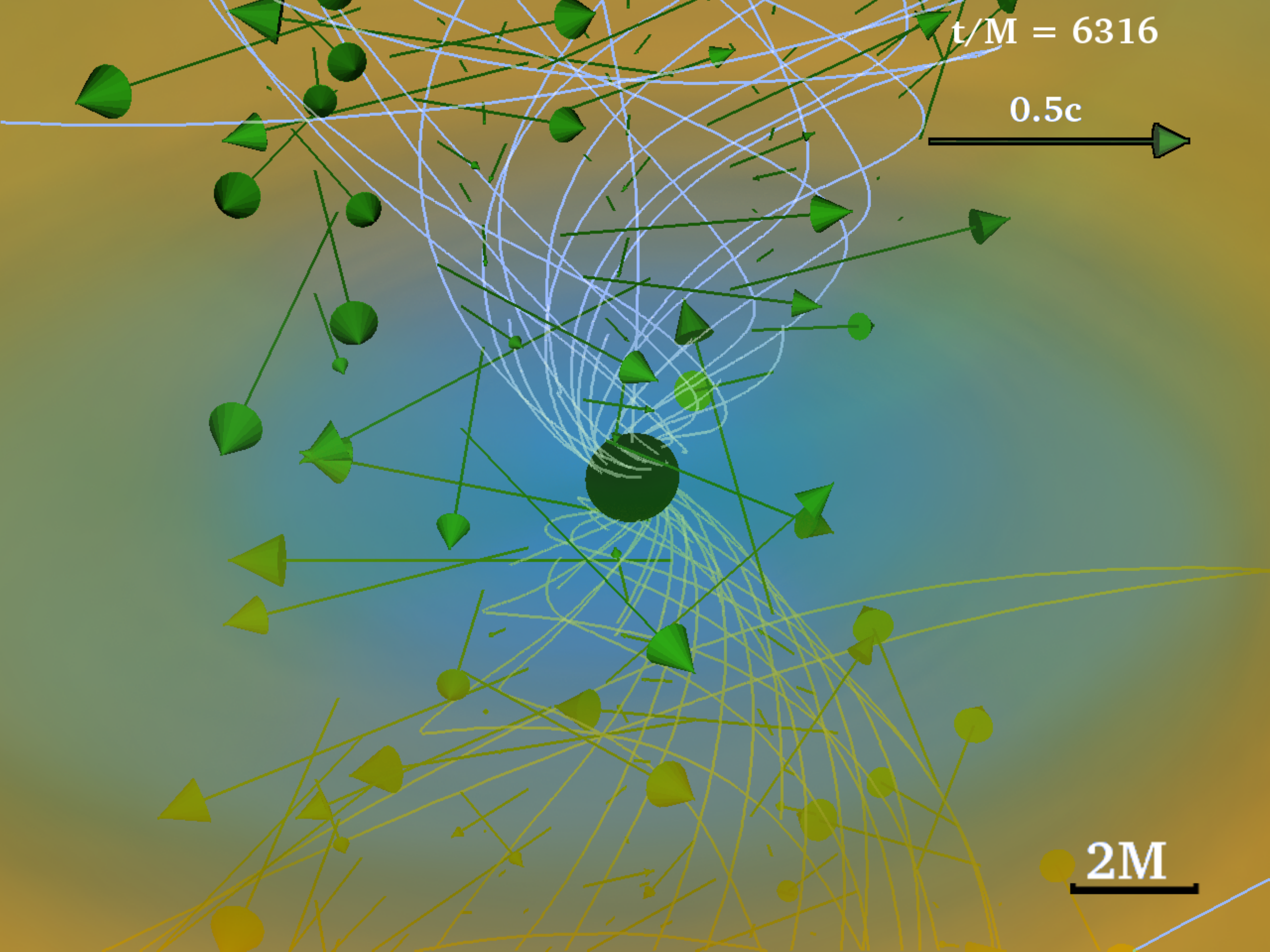}
  \caption{Snapshots of the rest-mass density, normalized to its
    initial maximum value~$\rho_{0,\rm max}=8.92\times
    10^{14}(1.4M_\odot/M_{\rm NS})^2\rm g\ cm^{-3}$ (log scale), at
    selected times. Arrows indicate plasma
    velocities and white lines show the magnetic field lines. Bottom
    panels highlight the system after an incipient jet is
    launched. Here,
    ${\rm M}=2.5\times 10^{-2}(M_{\rm NS}/1.4M_\odot)$ms$=7.58(M_{\rm
      NS}/1.4M_\odot)$km.
  \label{fig:Binit}}
\end{figure*}

In these early GRMHD BHNS studies, the B fields were confined to the NS
interior. However, NSs are expected to be endowed with dipole B fields
that extend into the NS exterior (as is required by pulsars).  A more
realistic initial configuration for a magnetized BHNS merger should
contain a NS endowed with a dipolar B field extending from the NS
interior well into the exterior. Two new features then arise: (1)
poloidal B-field lines attached to fluid elements thread the BH prior
to tidal disruption, and (2) following disruption, while the B field in
the disk remains predominantly toroidal, the initially poloidal
B field in the exterior maintains a strong poloidal component
threading the low-density debris (see Figure~\ref{fig:Binit}).

Motivated by these considerations, we perform ideal GRMHD simulations
of BHNS systems and show that they can launch incipient jets if the NS
is initially endowed with a dipolar B field extending into the
exterior. We use geometrized units,~$G=c=1$.

\setlength{\tabcolsep}{3.0pt}
\begin{deluxetable}{cccccc}
\scriptsize
\tablewidth{0pt}
\tablecaption{Summary of results \label{tab:models_BHNS}}
\tablecolumns{6}
\tablehead{
  \colhead{$\beta_0$} 
& \colhead{$\Gamma_L$ \tablenotemark{a}}  
& \colhead{$\dot{M}(M_\odot\ \rm s^{-1})$ \tablenotemark{b}}  
& \colhead{$M_{\rm disk}/M_{\rm NS}$}
& \colhead{$\alpha$-stress}
& \colhead{$L_{\rm EM}$ \tablenotemark{c}}
}
\startdata
0.01    &    1.25        &  0.25   & 9.9\% & 0.01-0.03   & 1.5$\times 10^{51}$  \\
0.05    &    1.3         &  0.39   & 9.1\% & 0.015-0.035 & 5.4$\times 10^{51}$  \\
0.1     &    1.2         &  0.31   & 9.4\% & 0.02-0.04   & 0.9$\times 10^{51}$  \\
\enddata
\tablenotetext{a}{Maximum fluid Lorentz factor near the end of the simulation.}
\tablenotetext{b}{Rest-mass accretion rate when the outflow reaches $100{\rm
    M}=758(M_{\rm NS}/1.4M_\odot)\rm km$ above the BH poles.}
\tablenotetext{c}{Outgoing Poynting luminosity in units of erg $s^{-1}$,
  time-averaged over the last $1000{\rm M}=25(M_{\rm NS}/1.4M_\odot)\rm ms$ of the evolution, after the
  jet is well-developed.}
\end{deluxetable}


\section{Methods}

We carry out the simulations employing the Illinois GRMHD
adaptive-mesh refinement code, which adopts the {\tt
  Cactus\footnote{http://www.cactuscode.org/}/Carpet\footnote{http://www.carpetcode.org/}}
infrastructure \citep{Carpet} and the {\tt AHFinderdirect} thorn
\citep{ahfinderdirect} to locate apparent horizons. This code has been
extensively tested~\citep{Etienne:2010ui} and used previously to study
different scenarios involving compact binaries and
B-fields~\citep{Etienne:2011ea,Etienne:2012te}. In all simulations, we
use 9 levels of refinement with two sets of nested refinement boxes
differing in size and resolution by factors of two.  One set is centered
on the NS and the other on the BH. The finest box around the BH (NS)
has a half-side length~$1.6\,R_{\rm BH}$ ($1.2\,R_{\rm
  NS}$). Here,~$R_{\rm BH}$~($R_{\rm NS}$)~is the initial BH (NS)
radius. The finest levels resolve the BH (NS) radius by 30 (40)
points. We set the outer boundary at~$200{\rm M}\simeq 1516(M_{\rm
  NS}/1.4M_{\odot})\,$km, and impose reflection symmetry across the
orbital plane.

The metric plus fluid initial data we use are identical to those in
case B of~\citet{Etienne:2012te}, and satisfy the conformal thin
sandwich equations~\citep[see, e.g.][]{Baumgarte10}. The BH:NS mass
ratio is $3:1$. While the likely BHNS binary mass ratios may be closer
to 7:1 \citep{Belczynskietal2010}, we choose 3:1 to compare with our
earlier studies. Note that remnant disks from 7:1 mass ratio BH--NS
mergers can have masses $\gtrsim 0.10M_\odot$, as obtained here,
provided the initial black hole spin parameter is $\gtrsim
0.8$~\citep{Foucart2012diskmasspred}. The initial NS is an
irrotational, unmagnetized, $\Gamma=2$ polytrope. Prior to tidal
disruption, the magnetic field will be simply advected with
(``frozen-into'') the fluid. To save computational resources and to
avoid buildup of numerical errors, we evolve the system until two
orbits prior to tidal disruption ($t=t_B$), at which point the NS is
seeded with a dynamically weak, dipolar B-field generated by a vector
potential~$A_\phi$ approximating the vector potential of a current
loop~\citep[see Eq. 2 in][]{Paschalidis:2013jsa}.  We choose the loop
current and radius such that in the interior the maximum value of the
ratio of magnetic to gas pressure is $\beta^{-1}=0.05$ which results
in an interior B-field strength ${B}_{int}\simeq
10^{17}(1.4M_\odot/M_{\rm NS})$G.  While the resulting B-field
strength is large, it is dynamically weak ($\beta^{-1}\ll 1$) and
enables us to provide an ``existence proof'' for jet launching with
the finite computational resources at our disposal. Specifically, we
show that a NS endowed with an initial dipolar magnetic field
extending from its interior into the exterior enables magnetic
launching of a jet following a BHNS merger. As this initial B-field is
dynamically unimportant in the NS interior, we expect that the
qualitative behavior obtained here will apply to other dynamically
weak field choices.

To evolve the exterior B field reliably and also mimic the force-free
conditions that likely characterize the exterior, at~$t=t_B$ we impose
a low but variable density atmosphere, as is typically done when
evolving exterior B fields with ideal MHD codes. In particular, we
construct an exterior in which the plasma parameter $\beta$ initially
is equal to some target value~$\beta_0 < 1$ everywhere. This choice
defines the NS surface as the place where the interior~$\beta$ falls
to~$\beta_{0}$. For~$\beta<\beta_{0}$ we are in the NS exterior and we
reset the low exterior rest-mass density
to~$\rho_0=\sqrt{0.5\,\beta_{0}\,\,b^2/K}$.  Here,~$b^2$ is the
magnetic energy density, and~$K=P/P_{\rm cold}$ is the exterior ratio
of gas pressure to cold pressure at~$t=t_B$.  The above prescription
guarantees a universal~$\beta=\beta_0$ in the exterior at $t=t_B$ and,
at the same time, captures one key aspect of force-free
electrodynamics, i.e., B-field pressure dominance. As the B-field
strength falls from the NS surface as~$1/r^3$, the above prescription
forces $\rho_0$ to fall as~$1/r^3$ as well. We note that our density
prescription in the exterior is not enforced throughout the
evolution. For the subsequent evolution, we evolve the density
everywhere according to the ideal GRMHD equations imposing a density
floor as is typically done in GRMHD codes. We vary
$\beta_{0}=0.1,\,0.05,\,0.01$ to study exterior conditions ranging
from partial to complete B-field pressure dominance, and check that
the outcome remains invariant.  For~$\beta_0= 0.01$ we have
$b^2/\rho_0\lesssim 1$ in the vicinity of the NS, and this case
provides our best approximation to a force-free environment. As long
as $b^2/\rho_0$ is not much larger than $1$, our
high-resolution-shock-capturing MHD code can handle the B-field
evolution~\citep[see][]{Duez:2005sf,Etienne:2010ui,Kiuchi2012}. With
our choice of~$\beta_{0}$ the amount of total rest-mass does not
increase by more than~$\sim 1\%$, even for~$\beta_0=0.1$. We also
evolve the initial data without seeding any B fields, as well as
repeating the calculation we performed in~\citet{Etienne:2011ea} with
initial poloidal B fields confined to the NS interior. We adopt
a~$\Gamma$-law EOS, allowing for shock heating.

All three~$\beta_0$ cases with an external B field yield a
magnetically driven, incipient jet, which is launched
at~$t_{\text{jet}}\sim 4000-6000{\rm M}\simeq 100-150(M_{\rm
  NS}/1.4M_\odot)$ms after tidal disruption (depending on the value
of~$\beta_0$). We define an incipient, magnetized jet as an unbound,
collimated, mildly relativistic outflow (Lorentz factor~$\sim 1.2$),
which is at least partially magnetically dominated. The outcome in all
three~$\beta_0$ cases satisfies the above definition. The time $t_{\rm
  jet}$ is also approximately the delay time between the peak GW
amplitude and the jet onset.


\section{Results}

As all of our cases demonstrate similar dynamics, we show snapshots
only for~$\beta_0=0.01$, the canonical case because the exterior
pressure is well dominated by the B field while~$\beta_0$ is within
the range of what our code can treat reliably. Results for all cases
are summarized in Table~\ref{tab:models_BHNS}.

\begin{figure}[t]
\centering
\includegraphics[width=0.40\textwidth,trim=0 0 0 0,clip=true]{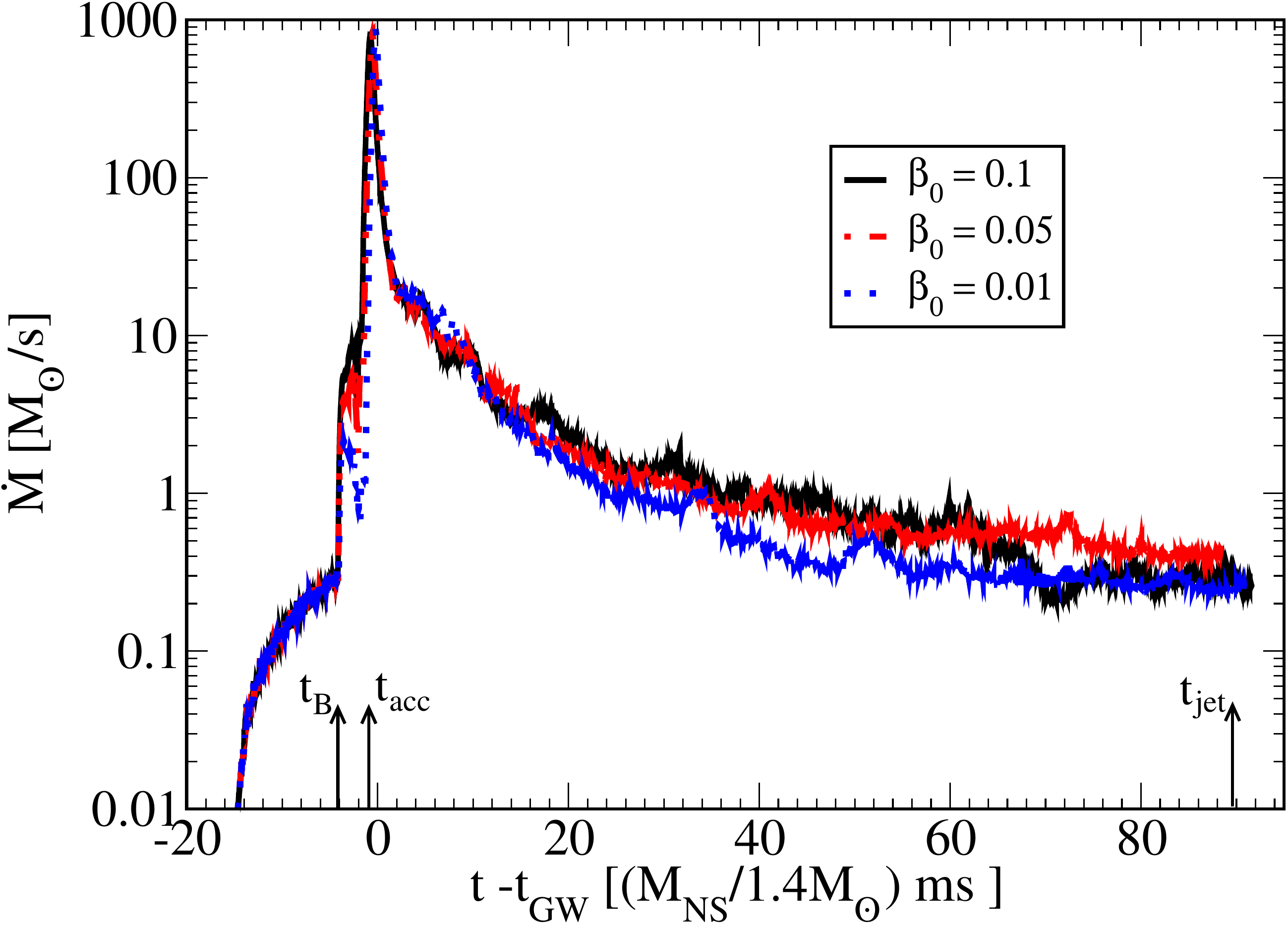}
\caption{Rest-mass accretion rates for all cases in
  Table~\ref{tab:models_BHNS}. Arrows indicate times $t_{\rm B}$,
  $t_{\rm acc}$, and $t_{\rm jet}$ for case $\beta_0=0.01$. Time is
  measured from the (retarded) time of the maximum GW amplitude, $t_{\rm
    GW}$.}
\label{fig:acretion_rate} 
\centering
\end{figure}

The merger and disk formation are displayed in Figure~\ref{fig:Binit}
and the dynamics are similar to what we reported in
\citet{Etienne:2011ea,Etienne:2012te} for case B, where the initial,
purely poloidal B fields were confined to the NS interior. Following
tidal disruption, a disk forms around a spinning BH with spin
parameter~$a/m\simeq 0.85$.  The initial dynamically weak, dipolar
B fields have little effect on the remnant disk rest-mass, which is
$\sim 15\%$ of~$M_{NS}$ at~$\sim 1000{\rm M}\simeq 25(M_{\rm
  NS}/1.4M_\odot)\rm ms$ after the time of peak accretion (at
$t=t_{\rm acc}=t_B+300{\rm M}$), as found in
\citet{UIUC_BHNS__BH_SPIN_PAPER}. The rest-mass accretion rate~($\dot
M$), which we compute via Eq. A11 of \citet{Farris:2009mt}, begins to
settle to an almost steady state at $t-t_{\rm acc}\sim 2000{\rm
  M}\simeq 50(M_{\rm NS}/1.4M_\odot)\rm ms$, and subsequently decays
slowly with time (see
Figure~\ref{fig:acretion_rate}). For~$\beta_0=0.01$, when the outflow
reaches~$100{\rm M}=758(M_{\rm NS}/1.4M_\odot)\rm km$ above the BH
poles, we find~$\dot M =0.25M_\odot\ \rm s^{-1}$ (see
Table~\ref{tab:models_BHNS} for the other cases), at which time the
disk mass is~$\sim 0.13M_\odot(M_{\rm NS}/1.4M_\odot)$. Thus, the disk
is expected to be accreted in~$\Delta t\sim M_{\rm disk}/{\dot M} \sim
0.5(M_{\rm NS}/1.4M_\odot)$s. It is interesting to note that the
engine's fuel -- the disk -- will be exhausted on a timescale entirely
consistent with the typical duration of sGRBs:~$T_{90}\sim
0.5s$~\citep[see, e.g.][]{Berger2014}, where~$T_{90}$ is the time over
which~$90\%$ of the total counts of gamma-rays have been detected.

While we resolve the wavelength of the fastest growing
magneto-rotational-instability (MRI) mode by at most five grid points,
we see evidence for turbulent B fields in meridional slices of the
disk. However, turbulence is not fully developed. Calculating the
effective Shakura--Sunyaev~$\alpha$ parameter associated with the
magnetic stresses~\citep[as defined in][]{PhysRevLett.109.221102}, we
find that in the innermost~$12{\rm M}\simeq 91(M_{\rm
  NS}/1.4M_\odot)$km of the disk and outside $\sim 5{\rm M}\simeq
38(M_{\rm NS}/1.4M_\odot)$km (the ISCO),~$\alpha$ is~$0.01-0.04$ (see
Table~\ref{tab:models_BHNS}), indicating that accretion is likely
driven by magnetic stresses. These values of the effective~$\alpha$
are similar to those found in other GRMHD simulations, including
rapidly spinning BHs~($a/m\sim 0.9$)~\citep{Krolik2007} such as ours.
Nevertheless,~$\alpha$ may depend on resolution \citep{Guan2009}:
higher resolution is required to accurately model the
magnetically driven turbulence and hence to determine the precise
lifetime of the remnant disk.

Neither the evolution without B field, nor the one with initial
B field confined in the interior launch jets or show any evidence for
an outflow. Instead, these runs exhibit inflows only, even when evolved
for~$5000{\rm M}\simeq 125(M_{\rm NS}/1.4M_\odot)\rm ms$.  

\begin{figure}
\centering
\includegraphics[width=0.34\textwidth,trim=0 140 0 0,clip=true]{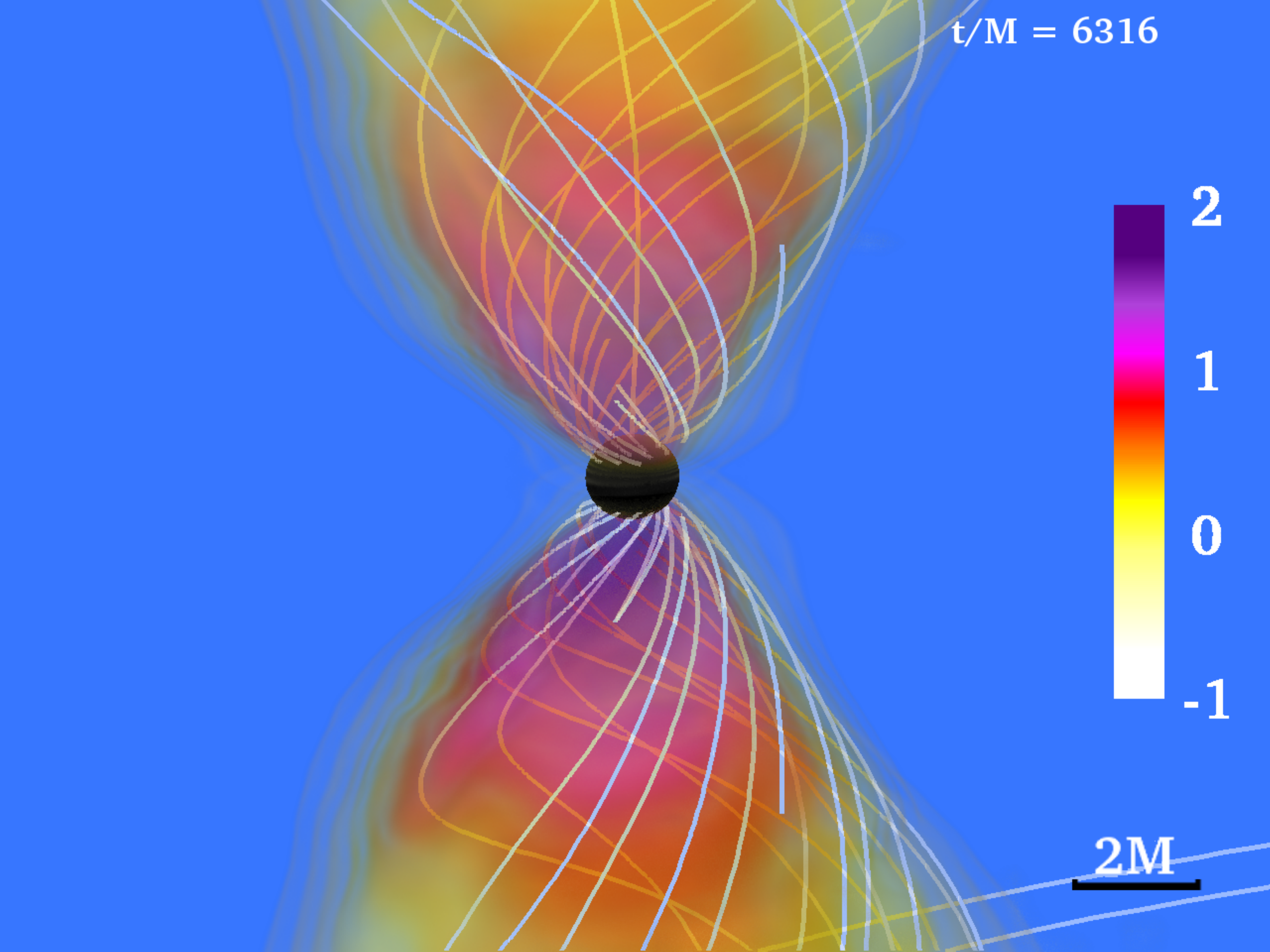}
  \caption{
    \label{fig:3D} 
    The ratio~$b^2/2\rho_0$ (log scale) at~$t-t_{\rm acc}=5316{\rm
      M}$. White lines indicate the B-field lines plotted in the
    funnel where~$b^2/2\rho_0 \geq 10^{-2}$. Magnetically-dominated
    areas~($b^2/\rho_0\geq 1$) extend to heights greater than~$15{\rm
      M}\simeq 25r_{\rm AH}$ above the BH horizon (shown as a black
    sphere). Here, $r_{\rm AH}=4.5(M_{NS}/1.4M_\odot)$km is the
    coordinate radius of the apparent horizon, which sets the length
    scale in the plot, and~${\rm M}=2.5\times 10^{-2}(M_{\rm
      NS}/1.4M_\odot)$ms$=7.58(M_{\rm NS}/1.4M_\odot)$km. }
\end{figure}

While a predominantly toroidal B field is contained in the remnant
disk, now the fluid elements inside the disk are linked via {\it
  exterior} B-field lines to other fluid elements in the disk and
fluid elements that were pushed far away during tidal disruption. As a
result, the B field outside the disk possesses a significant poloidal
component (see Figure~\ref{fig:Binit}).  The angular velocity on 2D
slices parallel to the orbital plane above the BH poles reveals that
the fluid is differentially rotating. Therefore, following tidal
disruption, magnetic winding of the plasma converts poloidal to
toroidal flux~\citep{Shapiro:2000zh}. This builds up the magnetic
field above the BH poles until the environment becomes force-free,
whereby the inflow is halted and eventually driven into an outflow
collimated by the B field (see Figure~\ref{fig:Binit}).  We find that
above the BH poles, the comoving B field is amplified
from~$\sim10^{13}(1.4M_\odot/M_{\rm NS})$G when the disk first settles
to~$\sim10^{15}(1.4M_\odot/M_{\rm NS})$G when the incipient jet is
launched.

In addition to differential rotation, the ``footpoints'' of the
B-field lines above the BH poles have an associated angular frequency
due to the Blandford-Znajek (BZ) effect \citep{Blandford1977} once the
environment becomes force-free~\citep{2004ApJ...611..977M}. As found
in \citet{Komissarov2001} for a monopole magnetic field around a BH
with spin parameter~$a/m=0.9$ (close to the value of~$0.85$ of the BH
remnant here), the ratio of the angular frequency of the B field
($\Omega_F=F_{t\theta}/F_{\theta\phi}$, where~$F_{\mu\nu}$ is the
Faraday tensor\footnote{This formula is strictly correct for
  spacetimes that are stationary and axisymmetric and expressed in
  Killing coordinates, which is approximately true in our simulations
  at late times.}) to the angular frequency of the BH
[$\Omega_H=(a/m)/2m\big(1+\sqrt{1-(a/m)^2}\big)$], increases
from~$\sim 0.49$ at the pole to~$\sim 0.53$ near the equator.
However, inside the jet funnel, where a force-free environment is
approached, the B-field geometry is approximately paraboloidal. Thus,
when the BZ effect operates, we expect that~$\Omega_F/\Omega_H \sim
0.4-0.5$ within the funnel as the angle from the BH spin axis varies
within 20 degrees \citep[see][]{Blandford1977,Yang2015}. Hence, the
B-field lines should rotate differentially. We
computed~$\Omega_F/\Omega_H$ on an x-z slice passing through the BH
centroid and along coordinate semicircles of radii~$r_{\rm AH}$ and
$2r_{\rm AH}$, where~$r_{\rm AH}$ is the BH apparent horizon radius,
and find that~$\Omega_F/\Omega_H\simeq 0.1-0.45$ within an opening
angle of~$20^\circ$ from the BH rotation axis. Therefore, the BZ
effect is likely operating, and the observed deviation
of~$\Omega_F/\Omega_H$ from~$\sim 0.4-0.5$ is probably due to
deviations from strict stationarity and axisymmetry, the gauge
required for our calculation of $\Omega_F$, and possibly inadequate
resolution. Values of $\Omega_F/\Omega_H\sim 0.45$ in the funnel have
been reported GRMHD accretion studies onto a single spinning BH with
$a/m=0.5$~\citep{2004ApJ...611..977M}.

Following jet launching, the outflowing fluid elements in the
asymptotically flat region have specific energy~$E=-u_0-1 > 0$; hence,
they are unbound. The characteristic maximum value of the Lorentz
factor reached in the outflow is~$\Gamma_L \sim 1.2$ (see
Table~\ref{tab:models_BHNS}). Thus, the incipient jet is only mildly
relativistic. However, in the nearly force-free areas in the funnel, we
find~$b^2/2\rho_0\sim 100$ (see Figure~\ref{fig:3D}). For
steady-state, axisymmetric Poynting-dominated jets, the maximum
attainable~$\Gamma_L$ in the (asymptotic) jet is equal to the
energy-to-mass flux ratio ($\simeq
b^2/2\rho_0$)~\citep{B2_over_2RHO_yields_target_Lorentz_factor}:
incipient jets thus can be accelerated to~$\Gamma_L\gtrsim 100$ as
required by sGRB models. While baryon loading will generally hinder
jet acceleration~\citep[see, e.g.,][]{2007ApJ...659..561M} and limit
the terminal~$\Gamma_L$, the BZ effect alone, which we capture, can
accelerate the jet to large~$\Gamma_L$ even in the presence of baryon
contamination~\citep{2005astro.ph..6368M}. Based on previous tests,
the increase in the magnetization in the funnel is robust, but values
of $b^2/2\rho_0$ exceeding $100$ may not be reliable.

The level of collimation in the outflow is measured by the funnel
opening angle. The boundary of the funnel area roughly coincides with
the $b^2/2\rho_0 \simeq 10^{-2}$ contour.  Based on this value, we
estimate the typical opening angle of the incipient jets in all cases
to be $\sim 20^\circ$ (see Figure~\ref{fig:3D}).

To eliminate the possibility that the plasma in the jets originates
from the artificial atmosphere only, we have tracked Lagrangian tracer
particles and have found that most of the plasma replenishing the
matter in the jet originates from the disk.

Computing the Poynting luminosity~$L_{\rm EM}$ on the surface of a
sphere of coordinate radius~$r=60{\rm M}\simeq 455(M_{\rm
  NS}/1.4M_\odot)\rm km$~\citep{Etienne:2011ea} we find~$L_{EM}\sim
10^{51}$ergs $s^{-1}$ (see Table~\ref{tab:models_BHNS}), i.e.,
consistent with characteristic sGRB luminosities
\citep{Berger2014}. This value is also consistent with the EM power
generated by the BZ effect: $L_{EM}\sim
10^{51}\,(a/m)^2\,(m/5.6M_\odot)^2\,(B/10^{15}\,G)^2 \rm erg\ s^{-1}$
\citep[see, e.g. Eq.~(4.50)~in][]{Thorne86}.

One of the main differences between the different $\beta_0$ cases is
the time at which the jet is launched. This is expected: the higher
$\beta_0$, the larger the inertia of the atmospheric matter with
respect to the B field, and the stronger the B field must become 
to overcome the ram pressure to launch a jet. Building up a
stronger B field by winding requires more time, hence higher $\beta_0$
generally implies a delay in jet launching, which is consistent with
our simulations.

\section{Conclusions}

We have shown that a NS in a BHNS system with an initially dynamically
weak, dipolar B field launches an incipient jet (an unbound,
collimated, mildly relativistic outflow) following tidal disruption.
The accretion timescale of the remnant disk and energy output are
consistent with those of typical sGRBs, demonstrating that the merger
of BHNSs can indeed be the engines powering sGRBs. Although our
results have been obtained with a high initial B field we expect that
a smaller initial field will yield the same qualitative outcome.  The
reason is the following: The B field is dynamically weak in the
stellar interior and does not affect the tidal disruption, formation
or structure of the disk. The amplification of the field following
disruption is largely due to magnetic winding and MRI. These should
combine to drive the field to values comparable to those found here
since amplification proceeds until appreciable differential rotational
and internal energy of the plasma, which is insensitive to the (weak)
initial field, is converted to magnetic energy. This amplification
should yield $B \sim 10^{15}$G at the BH poles nearly independent of
the initial NS B field, because BHNS simulations have shown that
characteristic ambient densities following tidal disruption are $\sim
10^9\rm gr\ cm^{-3}$ and magnetic launching is initiated when $b^2\sim
\rho_0 c^2$, yielding $b\sim 10^{15}$G. Also, the amplification in the
disk via MRI should not cease until the fields reach equipartition
values: typical pressures in these disks of $P\sim \rho_0 v^2\sim
10^{30}\rm dyn\ cm^{-2}$ should yield a magnetic field in the disk of
$b\sim 10^{15}$G. Winding occurs on an Alfven timescale, so
amplification may take longer the weaker the initial field. However, with
the fields reaching comparable values to the one found here, the
force-free exterior should drive a jet, and the BZ Poynting luminosity
should be comparable to the values in Table \ref{tab:models_BHNS}, in
agreement with observed sGRBs. We hope to investigate these issues and
confirm our expectations in future studies.

\acknowledgments
It is a pleasure to thank Zachariah Etienne, Charles Gammie, Roman
Gold, and James Stone for helpful discussions. We also thank the
Illinois Relativity group REU team (Sean E. Connelly, Abid Khan, and
Lingyi Kong) for assistance in creating Figures~\ref{fig:Binit}
and~\ref{fig:3D}, and R. Gold for sharing diagnostic 2D visualization
software that helped in our analysis. This work has been supported in
part by NSF grant PHY-1300903 and NASA grant NNX13AH44G at the
University of Illinois at Urbana-Champaign. V.P. gratefully acknowledges
support from a Fortner Fellowship at UIUC and support from the Simons
foundation and NSF grant PHY-1305682. This work used the Extreme
Science and Engineering Discovery Environment (XSEDE), which is
supported by NSF grant number OCI-1053575. This research is part of
the Blue Waters sustained-petascale computing project, which is
supported by the National Science Foundation (award number OCI
07-25070) and the state of Illinois. Blue Waters is a joint effort of
the University of Illinois at Urbana-Champaign and its National Center
for Supercomputing Applications.

\bibliographystyle{hapj}       
\bibliography{references}

\begin{thebibliography}{46}
\expandafter\ifx\csname natexlab\endcsname\relax\def\natexlab#1{#1}\fi

\bibitem[{{Acernese} \& {the VIRGO Collaboration}(2006)}]{VIRGO1}
{Acernese}, F., \& {the VIRGO Collaboration}. 2006, Class.~Quant.~Grav., 23,
  S635

\bibitem[{Baumgarte \& Shapiro(2010)}]{Baumgarte10}
Baumgarte, T., \& Shapiro, S. 2010, Numerical Relativity: Solving Einstein’s
  Equations on the Computer (Cambridge: Cambridge University Press)

\bibitem[{{Beckwith} {et~al.}(2008){Beckwith}, {Hawley}, \&
  {Krolik}}]{GRMHD_Jets_Req_Strong_Pol_fields}
{Beckwith}, K., {Hawley}, J.~F., \& {Krolik}, J.~H. 2008, \apj, 678, 1180

\bibitem[{{Belczynski} {et~al.}(2010){Belczynski}, {Dominik}, {Bulik},
  {O'Shaughnessy}, {Fryer}, \& {Holz}}]{Belczynskietal2010}
{Belczynski}, K., {Dominik}, M., {Bulik}, T., {O'Shaughnessy}, R., {Fryer}, C.,
  \& {Holz}, D.~E. 2010, \apjl, 715, L138, 1004.0386

\bibitem[{{Berger}(2014)}]{Berger2014}
{Berger}, E. 2014, Ann. Rev. Astron. Astroph., 52, 43

\bibitem[{{Blandford} \& {Znajek}(1977)}]{Blandford1977}
{Blandford}, R.~D., \& {Znajek}, R.~L. 1977, Mon. Not. Roy. Astron. Soc., 179,
  433

\bibitem[{Brown {et~al.}(2004)}]{LIGO2}
Brown, D.~A., {et~al.} 2004, Class.~Quant.~Grav., 21, S1625

\bibitem[{{Chawla} {et~al.}(2010){Chawla}, {Anderson}, {Besselman}, {Lehner},
  {Liebling}, {Motl}, \& {Neilsen}}]{cabllmn10}
{Chawla}, S., {Anderson}, M., {Besselman}, M., {Lehner}, L., {Liebling}, S.~L.,
  {Motl}, P.~M., \& {Neilsen}, D. 2010, Physical Review Letters, 105, 111101

\bibitem[{Deaton {et~al.}(2013)Deaton, Duez, Foucart, O'Connor, Ott,
  {et~al.}}]{Deaton:2013sla}
Deaton, M.~B., Duez, M.~D., Foucart, F., O'Connor, E., Ott, C.~D., {et~al.}
  2013, Astrophys.J., 776, 47

\bibitem[{Duez {et~al.}(2010)Duez, Foucart, Kidder, Ott, \&
  Teukolsky}]{Duez:2009yy}
Duez, M.~D., Foucart, F., Kidder, L.~E., Ott, C.~D., \& Teukolsky, S.~A. 2010,
  Class.Quant.Grav., 27, 114106

\bibitem[{Duez {et~al.}(2005)Duez, Liu, Shapiro, \& Stephens}]{Duez:2005sf}
Duez, M.~D., Liu, Y.~T., Shapiro, S.~L., \& Stephens, B.~C. 2005, Phys.Rev.,
  D72, 024028

\bibitem[{{East} {et~al.}(2012){East}, {Pretorius}, \& {Stephens}}]{East2012}
{East}, W.~E., {Pretorius}, F., \& {Stephens}, B.~C. 2012, \prd, 85, 124009

\bibitem[{Etienne {et~al.}(2012{\natexlab{a}})Etienne, Liu, Paschalidis, \&
  Shapiro}]{Etienne:2011ea}
Etienne, Z.~B., Liu, Y.~T., Paschalidis, V., \& Shapiro, S.~L.
  2012{\natexlab{a}}, Phys.Rev., D85, 064029

\bibitem[{Etienne {et~al.}(2010)Etienne, Liu, \& Shapiro}]{Etienne:2010ui}
Etienne, Z.~B., Liu, Y.~T., \& Shapiro, S.~L. 2010, Phys.Rev., D82, 084031

\bibitem[{{Etienne} {et~al.}(2009){Etienne}, {Liu}, {Shapiro}, \&
  {Baumgarte}}]{UIUC_BHNS__BH_SPIN_PAPER}
{Etienne}, Z.~B., {Liu}, Y.~T., {Shapiro}, S.~L., \& {Baumgarte}, T.~W. 2009,
  \prd, 79, 044024

\bibitem[{Etienne {et~al.}(2012{\natexlab{b}})Etienne, Paschalidis, \&
  Shapiro}]{Etienne:2012te}
Etienne, Z.~B., Paschalidis, V., \& Shapiro, S.~L. 2012{\natexlab{b}},
  Phys.Rev., D86, 084026

\bibitem[{Farris {et~al.}(2012)Farris, Gold, Paschalidis, Etienne, \&
  Shapiro}]{PhysRevLett.109.221102}
Farris, B.~D., Gold, R., Paschalidis, V., Etienne, Z.~B., \& Shapiro, S.~L.
  2012, Phys. Rev. Lett., 109, 221102

\bibitem[{Farris {et~al.}(2010)Farris, Liu, \& Shapiro}]{Farris:2009mt}
Farris, B.~D., Liu, Y.~T., \& Shapiro, S.~L. 2010, Phys.Rev., D81, 084008

\bibitem[{{Foucart}(2012)}]{Foucart2012diskmasspred}
{Foucart}, F. 2012, \prd, 86, 124007, 1207.6304

\bibitem[{Foucart {et~al.}(2014)Foucart, Deaton, Duez, O'Connor, Ott,
  {et~al.}}]{Foucart:2014nda}
Foucart, F., Deaton, M.~B., Duez, M.~D., O'Connor, E., Ott, C.~D., {et~al.}
  2014, Phys.Rev., D90, 024026

\bibitem[{{Guan} {et~al.}(2009){Guan}, {Gammie}, {Simon}, \&
  {Johnson}}]{Guan2009}
{Guan}, X., {Gammie}, C.~F., {Simon}, J.~B., \& {Johnson}, B.~M. 2009, Ap. J.,
  694, 1010

\bibitem[{Kiuchi {et~al.}(2014)Kiuchi, Kyutoku, Sekiguchi, Shibata, \&
  Wada}]{Kiuchi:2014hja}
Kiuchi, K., Kyutoku, K., Sekiguchi, Y., Shibata, M., \& Wada, T. 2014,
  Phys.Rev., D90, 041502, 1407.2660

\bibitem[{{Kiuchi} {et~al.}(2012){Kiuchi}, {Kyutoku}, \&
  {Shibata}}]{Kiuchi2012}
{Kiuchi}, K., {Kyutoku}, K., \& {Shibata}, M. 2012, \prd, 86, 064008

\bibitem[{{Komissarov}(2001)}]{Komissarov2001}
{Komissarov}, S.~S. 2001, Mon. Not. Roy. Astron. Soc., 326, L41

\bibitem[{{Krolik} \& {Hawley}(2007)}]{Krolik2007}
{Krolik}, J.~H., \& {Hawley}, J.~F. 2007, in American Institute of Physics
  Conference Series, Vol. 924, The Multicolored Landscape of Compact Objects
  and Their Explosive Origins, ed. T.~{di Salvo}, G.~L. {Israel},
  L.~{Piersant}, L.~{Burderi}, G.~{Matt}, A.~{Tornambe}, \& M.~T. {Menna},
  801--808, astro-ph/0611605

\bibitem[{{Kuroda} \& {LCGT Collaboration}(2010)}]{LCGT}
{Kuroda}, K., \& {LCGT Collaboration}. 2010, Classical and Quantum Gravity, 27,
  084004

\bibitem[{Kyutoku {et~al.}(2013)Kyutoku, Ioka, \& Shibata}]{Kyutoku:2013wxa}
Kyutoku, K., Ioka, K., \& Shibata, M. 2013, Phys.Rev., D88, 041503

\bibitem[{Lackey {et~al.}(2014)Lackey, Kyutoku, Shibata, Brady, \&
  Friedman}]{Lackey:2013axa}
Lackey, B.~D., Kyutoku, K., Shibata, M., Brady, P.~R., \& Friedman, J.~L. 2014,
  Phys.Rev., D89, 043009

\bibitem[{{Lee} \& {Ramirez-Ruiz}(2007)}]{LeeRamirezRuiz2007}
{Lee}, W.~H., \& {Ramirez-Ruiz}, E. 2007, New Journal of Physics, 9, 17,
  astro-ph/0701874

\bibitem[{Lovelace {et~al.}(2013)Lovelace, Duez, Foucart, Kidder, Pfeiffer,
  {et~al.}}]{Lovelace:2013vma}
Lovelace, G., Duez, M.~D., Foucart, F., Kidder, L.~E., Pfeiffer, H.~P.,
  {et~al.} 2013, Class.Quant.Grav., 30, 135004

\bibitem[{{L{\"u}ck} \& {the GEO600 collaboration}(2006)}]{GEO}
{L{\"u}ck}, H., \& {the GEO600 collaboration}. 2006, Class.~Quant.~Grav., 23,
  S71

\bibitem[{{McKinney}(2005)}]{2005astro.ph..6368M}
{McKinney}, J.~C. 2005, ArXiv Astrophysics e-prints, arXiv:astro-ph/0506368

\bibitem[{{McKinney} \& {Gammie}(2004)}]{2004ApJ...611..977M}
{McKinney}, J.~C., \& {Gammie}, C.~F. 2004, \apj, 611, 977

\bibitem[{Meszaros(2006)}]{Meszaros:2006rc}
Meszaros, P. 2006, Rept.Prog.Phys., 69, 2259

\bibitem[{{Metzger} {et~al.}(2007){Metzger}, {Thompson}, \&
  {Quataert}}]{2007ApJ...659..561M}
{Metzger}, B.~D., {Thompson}, T.~A., \& {Quataert}, E. 2007, \apj, 659, 561

\bibitem[{Pannarale {et~al.}(2013)Pannarale, Berti, Kyutoku, \&
  Shibata}]{Pannarale:2013uoa}
Pannarale, F., Berti, E., Kyutoku, K., \& Shibata, M. 2013, Phys.Rev., D88,
  084011

\bibitem[{Paschalidis {et~al.}(2013)Paschalidis, Etienne, \&
  Shapiro}]{Paschalidis:2013jsa}
Paschalidis, V., Etienne, Z.~B., \& Shapiro, S.~L. 2013, Phys.Rev., D88, 021504

\bibitem[{{Rezzolla} {et~al.}(2011){Rezzolla}, {Giacomazzo}, {Baiotti},
  {Granot}, {Kouveliotou}, \& {Aloy}}]{ML2011}
{Rezzolla}, L., {Giacomazzo}, B., {Baiotti}, L., {Granot}, J., {Kouveliotou},
  C., \& {Aloy}, M.~A. 2011, \apjl, 732, L6

\bibitem[{Schnetter {et~al.}(2004)Schnetter, Hawley, \& Hawke}]{Carpet}
Schnetter, E., Hawley, S.~H., \& Hawke, I. 2004, Class. Quantum Grav., 21,
  1465, arXiv:gr-qc/0310042

\bibitem[{Shapiro(2000)}]{Shapiro:2000zh}
Shapiro, S.~L. 2000, Astrophys.J., 544, 397

\bibitem[{{Shibata} \& {Taniguchi}(2011)}]{st11}
{Shibata}, M., \& {Taniguchi}, K. 2011, Living Reviews in Relativity, 14, 6

\bibitem[{Tanaka {et~al.}(2014)Tanaka, Hotokezaka, Kyutoku, Wanajo, Kiuchi,
  {et~al.}}]{Tanaka:2013ixa}
Tanaka, M., Hotokezaka, K., Kyutoku, K., Wanajo, S., Kiuchi, K., {et~al.} 2014,
  Astrophys.J., 780, 31

\bibitem[{{Thornburg}(2004)}]{ahfinderdirect}
{Thornburg}, J. 2004, Class.~Quant.~Grav., 21, 743

\bibitem[{Thorne {et~al.}(1986)Thorne, Price, \& Macdonald}]{Thorne86}
Thorne, K.~S., Price, R.~H., \& Macdonald, D.~A. 1986, The Membrane Paradigm
  (New Haven: Yale University Press)

\bibitem[{{Vlahakis} \&
  {K{\"o}nigl}(2003)}]{B2_over_2RHO_yields_target_Lorentz_factor}
{Vlahakis}, N., \& {K{\"o}nigl}, A. 2003, \apj, 596, 1080

\bibitem[{{Yang} {et~al.}(2015){Yang}, {Zhang}, \& {Lehner}}]{Yang2015}
{Yang}, H., {Zhang}, F., \& {Lehner}, L. 2015, ArXiv e-prints, 1503.06788

\end{thebibliography}

\end{document}